\documentclass[prd,letterpaper,twocolumn,superscriptaddress,nofootinbib]{revtex4}
\usepackage{amsmath,amssymb,epsfig,natbib,bm,psfrag}

\begin{document}

\renewcommand{\tilde}{\widetilde}

\newcommand{\HALOFIT}{\textsc{halofit}}
\newcommand{\Mpc}{\text{Mpc}}
\newcommand{\half}{{\textstyle \frac{1}{2}}}
\newcommand{\third}{{\textstyle \frac{1}{3}}}
\newcommand{\numfrac}[2]{{\textstyle \frac{#1}{#2}}}
\newcommand{\ra}{\rangle}
\newcommand{\la}{\langle}
\renewcommand{\d}{\text{d}}
\newcommand{\grad}{\mbox{\boldmath$\nabla$}}

\newcommand{\begm}{\begin{pmatrix}}
\newcommand{\enm}{\end{pmatrix}}

\newcommand{\threej}[6]{{\begm #1 & #2 & #3 \\ #4 & #5 & #6 \enm}}
\newcommand{\fsky}{f_{\text{sky}}}

\newcommand{\cla}{\mathcal{A}}
\newcommand{\clb}{\mathcal{B}}
\newcommand{\clc}{\mathcal{C}}
\newcommand{\cle}{\mathcal{E}}
\newcommand{\clf}{\mathcal{F}}
\newcommand{\clg}{\mathcal{G}}
\newcommand{\clh}{\mathcal{H}}
\newcommand{\cli}{\mathcal{I}}
\newcommand{\clj}{\mathcal{J}}
\newcommand{\clk}{\mathcal{K}}
\newcommand{\cll}{\mathcal{L}}
\newcommand{\clm}{\mathcal{M}}
\newcommand{\cln}{\mathcal{N}}
\newcommand{\clo}{\mathcal{O}}
\newcommand{\clp}{\mathcal{P}}
\newcommand{\clq}{\mathcal{Q}}
\newcommand{\clr}{\mathcal{R}}
\newcommand{\cls}{\mathcal{S}}
\newcommand{\clt}{\mathcal{T}}
\newcommand{\clu}{\mathcal{U}}
\newcommand{\clv}{\mathcal{V}}
\newcommand{\clw}{\mathcal{W}}
\newcommand{\clx}{\mathcal{X}}
\newcommand{\cly}{\mathcal{Y}}
\newcommand{\clz}{\mathcal{Z}}
% Miscellaneous
\newcommand{\CMBFAST}{\textsc{cmbfast}}
\newcommand{\CAMB}{\textsc{camb}}
\newcommand{\COSMOMC}{\textsc{CosmoMC}}
\newcommand{\Healpix}{\textsc{healpix}}
\newcommand{\IGLOO}{\textsc{igloo}}
\newcommand{\GLESP}{\textsc{glesp}}

\newcommand{\Omtot}{\Omega_{\mathrm{tot}}}
\newcommand{\Omb}{\Omega_{\mathrm{b}}}
\newcommand{\Omc}{\Omega_{\mathrm{c}}}
\newcommand{\Omm}{\Omega_{\mathrm{m}}}
\newcommand{\omb}{\omega_{\mathrm{b}}}
\newcommand{\omc}{\omega_{\mathrm{c}}}
\newcommand{\omm}{\omega_{\mathrm{m}}}
\newcommand{\Omdm}{\Omega_{\mathrm{DM}}}
\newcommand{\Omnu}{\Omega_{\nu}}

\newcommand{\Oml}{\Omega_\Lambda}
\newcommand{\OmK}{\Omega_K}

\newcommand{\Hunit}{~\text{km}~\text{s}^{-1} \Mpc^{-1}}
\newcommand{\Gyr}{{\rm Gyr}}

\newcommand{\nrun}{n_{\text{run}}}

\newcommand{\lmax}{l_{\text{max}}}

\newcommand{\zre}{z_{\text{re}}}
\newcommand{\mpl}{m_{\text{Pl}}}

\newcommand{\vphi}{\mathbf{\psi}}
\newcommand{\vv}{\mathbf{v}}
\newcommand{\vd}{\mathbf{d}}
\newcommand{\vC}{\mathbf{C}}
\newcommand{\vT}{\mathbf{T}}
\newcommand{\vX}{\mathbf{X}}
\newcommand{\vn}{\hat{\mathbf{n}}}
\newcommand{\vy}{\mathbf{y}}
\newcommand{\mN}{\bm{N}}
\newcommand{\eV}{\,\text{eV}}
\newcommand{\vtheta}{\bm{\theta}}
\newcommand{\tT}{\tilde{T}}
\newcommand{\tE}{\tilde{E}}
\newcommand{\tB}{\tilde{B}}

\newcommand{\mCh}{\hat{\bm{C}}}
\newcommand{\Ch}{\hat{C}}

\newcommand{\Bt}{\tilde{B}}
\newcommand{\Et}{\tilde{E}}
\newcommand{\bld}[1]{\mathrm{#1}}
\newcommand{\mLambda}{\bm{\Lambda}}
\newcommand{\mA}{\bm{A}}
\newcommand{\mC}{\bm{C}}
\newcommand{\mQ}{\bm{Q}}
\newcommand{\mU}{\bm{U}}
\newcommand{\mX}{\bm{X}}
\newcommand{\mV}{\bm{V}}
\newcommand{\mP}{\bm{P}}
\newcommand{\mR}{\bm{R}}
\newcommand{\mW}{\bm{W}}
\newcommand{\mD}{\bm{D}}
\newcommand{\mI}{\bm{I}}
\newcommand{\mH}{\bm{H}}
\newcommand{\mM}{\bm{M}}
\newcommand{\mS}{\bm{S}}
\newcommand{\mzero}{\bm{0}}
\newcommand{\mL}{\bm{L}}

\newcommand{\btheta}{\bm{\theta}}
\newcommand{\bphi}{\bm{\psi}}

\newcommand{\vb}{\mathbf{b}}
\newcommand{\vA}{\mathbf{A}}
\newcommand{\vAt}{\tilde{\mathbf{A}}}
\newcommand{\ve}{\mathbf{e}}
\newcommand{\vE}{\mathbf{E}}
\newcommand{\vB}{\mathbf{B}}
\newcommand{\vEt}{\tilde{\mathbf{E}}}
\newcommand{\vBt}{\tilde{\mathbf{B}}}
\newcommand{\vEw}{\mathbf{E}_W}
\newcommand{\vBw}{\mathbf{B}_W}
\newcommand{\vx}{\mathbf{x}}
\newcommand{\vXt}{\tilde{\vX}}
\newcommand{\vXb}{\bar{\vX}}
\newcommand{\vTb}{\bar{\vT}}
\newcommand{\vTt}{\tilde{\vT}}
\newcommand{\vY}{\mathbf{Y}}
\newcommand{\vBwr}{{\vBw^{(R)}}}
\newcommand{\RW}{{W^{(R)}}}

\newcommand{\mUt}{\tilde{\mU}}
\newcommand{\mVt}{\tilde{\mV}}
\newcommand{\mDt}{\tilde{\mD}}

\newcommand{\Rot}{\begm \mzero &\mI \\ -\mI & \mzero \enm}
\newcommand{\Pt}{\begm \vEt \\ \vBt \enm}

\newcommand{\edth}{\,\eth\,}
\renewcommand{\beth}{\,\overline{\eth}\,}

\newcommand{\sE}{{}_{|s|}E}
\newcommand{\sB}{{}_{|s|}B}
\newcommand{\sElm}{\sE_{lm}}
\newcommand{\sBlm}{\sB_{lm}}

%%%%%%%%%%%%%%%%%%%%%%%%%%%%%%%%%%%%%%%%%%%%%%%%%%%%%%%%%%%%%%%%%%
%                       Title matter                             %
%%%%%%%%%%%%%%%%%%%%%%%%%%%%%%%%%%%%%%%%%%%%%%%%%%%%%%%%%%%%%%%%%%

% title and affiliations
\title{Lensed CMB simulation and parameter estimation}

\author{Antony Lewis}
 \homepage{http://cosmologist.info}
 \affiliation{CITA, 60 St. George St, Toronto M5S 3H8, ON, Canada.}

\begin{abstract}

Modelling of the weak lensing of the CMB will be crucial to obtain
correct cosmological parameter constraints from forthcoming precision CMB anisotropy observations.  The lensing affects the power spectrum as well
as inducing non-Gaussianities. We discuss the simulation of full sky CMB maps in the weak
lensing approximation and describe a fast numerical code.
The series expansion in the deflection angle cannot be used to
simulate accurate CMB maps, so a pixel remapping must be used.
For parameter estimation accounting for the change
in the power spectrum but assuming Gaussianity is sufficient to obtain
accurate results up to Planck sensitivity using current tools. A fuller
analysis may be required to obtain accurate error estimates and for more sensitive observations.
We demonstrate a simple full sky
simulation and subsequent parameter estimation at Planck-like
sensitivity. The lensed CMB simulation and parameter estimation codes
are publicly available.

\vspace{\baselineskip}
\end{abstract}


\maketitle

%%%%%%%%%%%%%%%%%%%%%%%%%%%%%%%%%%%%%%%%%%%%%%%%%%%%%%%%%%%%%%%%%%
%                         Main body                              %
%%%%%%%%%%%%%%%%%%%%%%%%%%%%%%%%%%%%%%%%%%%%%%%%%%%%%%%%%%%%%%%%%%
\section{Introduction}

The CMB temperature and polarization anisotropies are being measured
with ever more precision. The statistics of the anisotropies can
provide valuable limits on cosmological parameters as well as
constrain early universe physics. As we enter the era of precision
measurement with signal-dominated observations out to small scales,
the non-linear effects will become important.
One of the most important of these on scales of most interest
for parameter estimation is that of weak lensing. Fortunately it can be
modelled accurately using linear physics: the linear potentials
along the line sight lensing the linear perturbations at the last
scattering surface~\cite{Seljak:1996ve,Zaldarriaga:1998ar,Hu:2000ee}.
Modelling of fully non-linear evolution is not required for the near
future on
scales with $l \alt 2000$, and non-linear corrections can be applied to
the lensing potential if and when required.
On smaller scales the situation
becomes much more complicated anyway due to point sources, beam size
and other non-linear effects.

Lensing induces non-Gaussianities in the lensed CMB sky, and also
changes the power spectra of the perturbations.
Lensing will start to have an observable effect on the power spectrum
very shortly, and hence needs to be taken into account to obtain correct
parameter constraints and error bars. For
future observations, including the
Planck\footnote{\url{http://sci.esa.int/planck}} satellite and forthcoming
ground based telescopes the effect is very
important. In this paper we describe the simulation of lensed CMB
maps (including the full non-Gaussian structure), but show that using an accurate calculation of the lensed CMB
power spectra~\cite{Challinor:2005jy} a naive parameter
estimation (neglecting non-Gaussianities) works rather well up to
Planck
sensitivities.
Observations at higher sensitivities and resolutions will probably require a
fuller analysis accounting for the full non-Gaussian distribution of
the lensed sky, an important problem that we do not tackle here. Our
simulation code can be used for testing future methods, and the simple
power spectrum parameter estimation method can act as a useful
baseline for future improvements.

\section{Weak lensing of the CMB}

The small scale CMB anisotropy is dominated by the emission from the
last scattering surface at redshift $z\sim 1000$.
Weak lensing of the CMB deflects photons coming from an original direction
$\vn'$ on the last scattering surface to an observed direction $\vn$
on the sky today, so a lensed CMB field is given by
\begin{equation}
 \tilde{X}(\vn) = X(\vn')
\end{equation}
where $X$ is the unlensed field. The arcminute-scale displacement of the points is
determined by the potential along the line of sight to the last
scattering surface, conveniently
encapsulated into an integrated lensing potential $\psi$~\cite{Hu:2000ee}.
The deflection vector is given by the gradient of the lensing
potential  $\grad\psi(\vn)$, where
 $\grad$ is the covariant derivative on the sphere.
The vector $\vn'$ is obtained from $\vn$ by moving its end on
the surface of a unit sphere a distance $|\grad\psi(\vn)|$ along a geodesic
 in the direction of
 $\grad\psi(\vn)$~\cite{Hu:2000ee,Challinor02}. This is
 sometimes written as $\vn' = \vn + \grad\psi(\vn)$.
To our level of approximation $|\grad\psi|$ is assumed to be constant between $\vn$ and $\vn'$,
consistent with working out the lensing potential in the Born
approximation (evaluating the potential along the undeflected
path). Lensing deflections are a few arcminutes, but are coherent over
degree scales, so this is a good approximation.

\begin{figure*}
\begin{center}
\psfig{figure=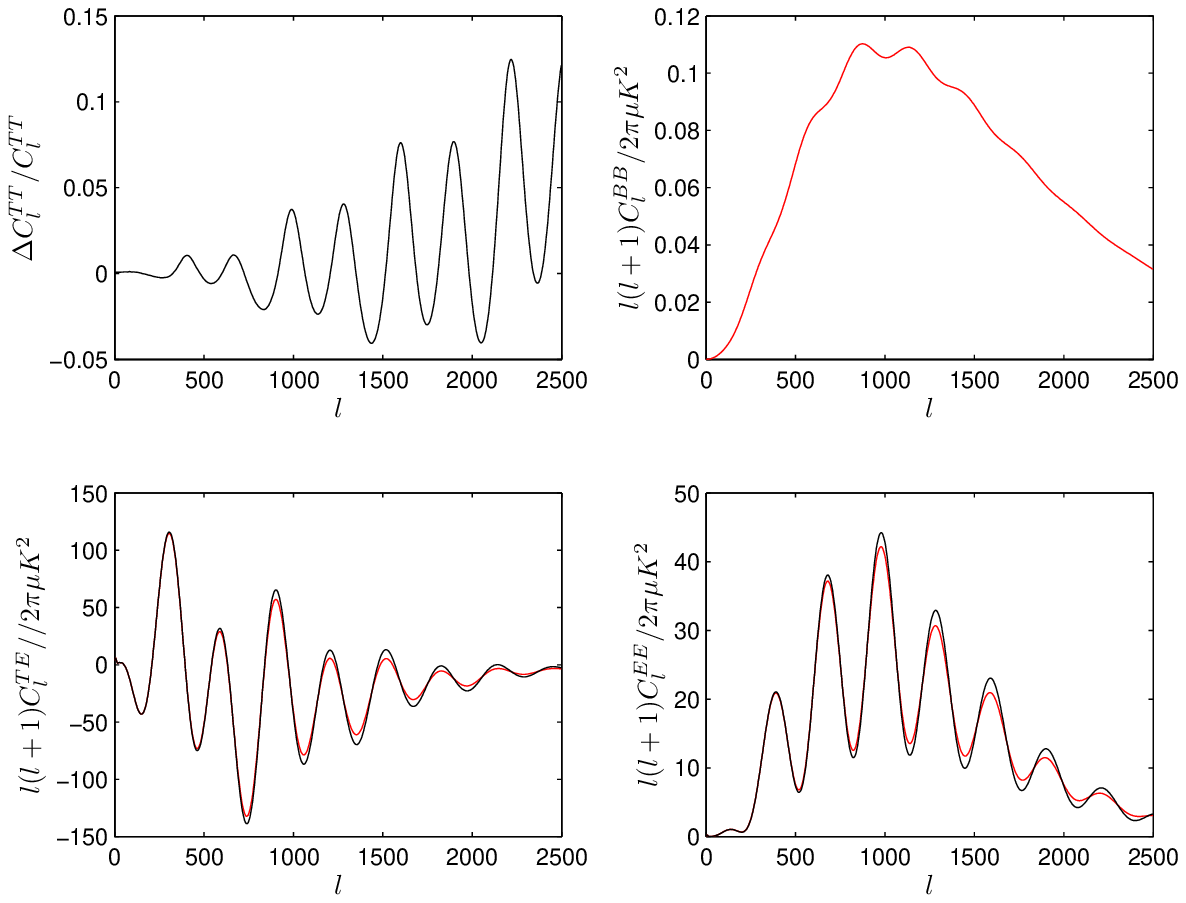,angle=0,width=16cm,clip=}
\caption{The effect of lensing on the CMB power spectra. The top plots
  show the fractional change in the temperature spectrum $C_l^{TT}$
  and the lensing-induced B-polarization spectrum $C_l^{BB}$. The
  bottom plots show the lensed (grey/red, less peaked) and unlensed (black)
  T-E cross-correlation $C_l^{TE}$ and E-polarization $C_l^{EE}$ power
  spectra.  All results are for the fiducial model given in the text, and the lensed
  B-mode power spectrum shown is not very accurate due to the neglect of
  non-linear evolution in the lensing potential.
\label{lensedcls}}
\end{center}
\end{figure*}

The power spectrum for the lensing potential $C^{\psi\psi}_l$ (and the
correlation to the temperature $C^{\psi T}_l$) can be computed
numerically in linear theory for a particular cosmological model using
\CAMB\footnote{\url{http://camb.info}}~\cite{Lewis:1999bs,Hu00,Challinor:2005jy}.
From the power spectrum simulated realizations can be made
assuming Gaussianity of the primordial fields.  Non-linear evolution
of the potential changed the power spectrum $C^{\psi\psi}_l$ on small
scales and also makes
the $\psi$ distribution somewhat non-Gaussian. The power spectrum $l^4
C^{\psi\psi}_l$ peaks at $l\sim 60$, however small scales
where non-linear evolution is important at late times only contribute
to $\psi$ at $l\gg 60$. On these scales the
contributions to the lensing potential come from a rather broad
range of redshifts from $1\alt z \alt 10$, so even at $l\sim 1000$
the late time non-linear evolution does not radically change
$C^{\psi\psi}_l$.
We therefore neglect the small effect of
non-linear evolution here, though it can become important for high
resolution polarization $B$-mode experiments. Using
\CAMB\ with \HALOFIT~\cite{Smith:2002dz,Challinor:2005jy} the
non-linear evolution of $\psi$ can be estimated to change the lensed
temperature power spectrum
$\tilde{C}_l^{TT}$ by about $\sim 0.2\%$ at $l\sim 2000$, though growing to
$1\%$ or more on smaller scales.

The significant effect of lensing on the CMB power spectra is shown in
Fig.~\ref{lensedcls}, where the lensed power spectra are computed
accurately numerically using the correlation function method of
Ref.~\cite{Challinor:2005jy}. The lensing smoothes out features in the
temperature and polarization power spectra, changing the $C_l$ peaks
by up to $20\%$ for the E-polarization on the scales of interest.
 Weak lensing does not change the total variance of the CMB anisotropies, with
\begin{equation}
\sum_l (2l+1) \tilde{C}_l = \sum_l (2l+1) C_l,
\end{equation}
and similarly for the polarization~\cite{Challinor:2005jy}.
This encapsulates that fact that the observation in any fixed
direction is just a displaced view of the last scattering surface, and
hence is Gaussian and with the same variance as if there were no
lensing. Lensing induces a non-Gaussian spatial correlation structure
to the lensed CMB fields but does not alter the variance at a point.

We now move on to discuss how to simulate full-sky lensed CMB maps.

\subsection{Series expansion}
\begin{figure}
\begin{center}
\psfig{figure=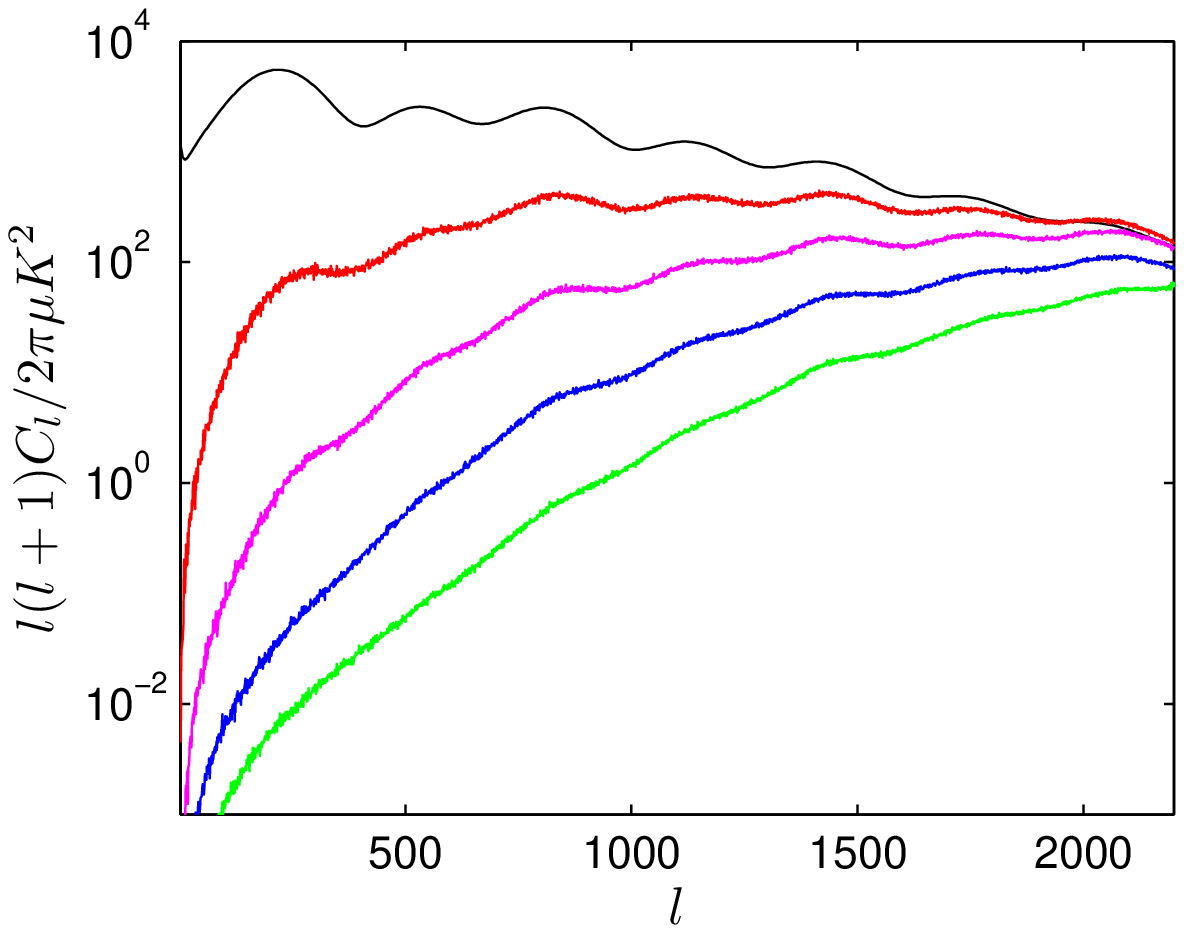,angle=0,width=8.5cm}
\caption{Power spectra from realizations of the 1st, 2nd, 3rd and 4th order terms in the
  lensing potential series expansion of the lensed temperature $\tilde{T}_{lm}$
 compared to the full lensed $\tilde{C}_l$ (top). The spectra may
 contain some pixelization error.
\label{series_contribs}}
\end{center}
\end{figure}

A common procedure for working with CMB weak lensing it to perform a
series expansion in the deflection angle, so for the temperature
\begin{eqnarray}
\tilde{T}(\vn) &=& T(\vn + \grad\psi)\nonumber\\
&=& T + \grad_a T \grad^a\psi + \half \grad^a\vphi
\grad^b \vphi  \grad_a\grad_b T  + \dots\,
\end{eqnarray}
and the last line is evaluated at $\vn$.
This expansion will be valid for scales much smaller than the deflection
angle. On smaller scales the unlensed fields are deflected by a
distance comparable to their wavelength, and the change in phase is not
a small perturbation.
%More quantitatively, considering the
%temperature,
%one can show that averaged over lensing potential realizations
%\begin{equation}
%\la \tilde{T}_{lm} \ra_\psi = e^{-l(l+1)R_\psi/2} T_{lm}
%\end{equation}
%where
%\begin{equation}
%R_\psi \equiv \frac{1}{2} \sum_l l(l+1) \frac{2l+1}{4\pi}
%C_l^{\psi\psi} \sim 3\times 10^{-7}.
%\end{equation}
%A minimum requirement for the
%validity of the series expansion is therefore that $l(l+1)R_\psi/2 <
%1$ (i.e. $l \alt 2500$). The expansion is therefore only likely to be
%accurate for $l \ll 2500$.
Even on scales where the expansion is valid, the expansion only converges
relatively slowly. The power spectra of the individual terms in the
expansion are shown
in Fig.~\ref{series_contribs} up to fourth order. This clearly shows that a low order series expansion cannot be used for accurate
map simulation due to the large variance of higher order terms on
small scales. The lowest order series expansion may be a useful approximation for
some applications (e.g. Ref.~\cite{Okamoto03}), but can easily give
results which are sufficiently inaccurate to be problematic. For
example it was shown in Ref.~\cite{Challinor:2005jy} that the lensed CMB
power spectrum computed from the lowest order series expansion of Ref.~\cite{Hu00} gives
lensing corrections which are incorrect by an order unity factor on
small scales.

\subsection{Re-mapping points}

Since the series expansion is not accurate enough for making simulated
maps at high resolution, the best way to proceed is to re-map the
anisotropy field by the
deflection vector $\grad\psi$ as a function of position. The deflection
field is a vector field on the sphere, and can be easily simulated
using spin-1 spherical harmonics (or, equivalently, vector spherical
harmonics). Full details of the simulation process for the temperature
and polarization on the sphere are
given in the Appendix, along with details of a fast multi-processor
implementation using a modified version of
\Healpix\footnote{\url{http://www.eso.org/science/healpix/}} 1.2~\cite{Gorski:2004by}.

It is important to remember that the lensing deflects the physical
anisotropy field, not the field after beam and pixel convolution. For
this reason sky simulation and re-mapping have to be performed at high resolution
regardless of how broad the observational beam is. The lensed field
can then be convolved as required.

\section{Parameter estimation}

With future precision observations we would ideally like to extract information about
both the unlensed CMB (which has simple Gaussian statistical
properties), and about the lensing
potential (which contains additional information).
Gravitational lensing of scalar E-mode polarization may ultimately dominate the
B-mode polarization signal from primordial gravitational waves,
so an accurate treatment of the lensing will become
essential~\cite{Hirata:2003ka,Knox02,Kesden:2002ku}. In the more immediate future
gravitational lensing will have a significant effect on the statistics
of the observed CMB, and must be accounted for to obtain reliable
parameter estimates. The non-Gaussianities induced by lensing must
also be accounted for when
attempting to assess the degree of primordial non-Gaussianity (an
important probe of early universe physics).

The amount of information that can be learned about the lensing
potential from the non-Gaussianities depends on the noise
level~\cite{Kesden:2003cc,Okamoto03,Hirata:2002jy}. For observations up to Planck
sensitivity any reconstructed map of the lensing potential would be
completely noise dominated, and the extra information this contains is
rather limited. However the effect of lensing on the power spectrum is
significant and certainly cannot be ignored. In this paper we do not
address the problem of handling the full likelihood function, but
merely show that using the lensed power spectrum and approximating the
lensed field as Gaussian is sufficient to obtain good parameter
estimates at Planck sensitivity. More general methods for handling the
lensing likelihood function will be discussed in a future paper if they can be made to work.

\subsection{Gaussian $C_l$ Likelihood function}

Spherical harmonic coefficients of a Gaussian temperature and polarization field
on the sky can be used to define estimators of the covariance
\begin{equation}
\Ch_l^{WX} = \frac{1}{2l+1} \sum_m  W_{lm}^* X_{lm}
\end{equation}
where $W_{lm}$ and $X_{lm}$ are spherical harmonic coefficients of the temperature $T$, E-polarization $E$ or
B-polarization $B$. These are unbiased in that averaged over
realizations $\la \Ch_l^{WX} \ra = C_l^{WX}$.

The assumed Gaussianity of $T_{lm}$, $E_{lm}$ and $B_{lm}$ gives the
following full-sky likelihood function:
\begin{equation}
-2\log P(\mCh_l|\mC_l) = (2l+1) \left\{\text{Tr}\left[ \mCh_l \mC^{-1}_l
  \right] + \log |\mC_l| \right\}
\end{equation}
(to within an irrelevant constant) where
\begin{equation}
\mC_l =  \begm C^{TT}_l & C^{TE}_l & 0 \\ C^{TE}_l & C^{EE}_l & 0 \\ 0 & 0 &
C^{BB}_l \enm
\end{equation}
and $\mCh$ is the corresponding matrix of estimators. In the presence
of instrumental noise the $C_l$ and $\Ch_l$ should include the noise variance.
We have assumed a statistically
parity invariant ensemble so that $C^{BT} = C^{BE} = 0$.

For the lensed sky this is not the correct relation because the lensed
sky is not Gaussian if the lensing potential is not fixed.
However as discussed below, replacing $C_l$ by $\tilde{C}_l$ does not give significantly biased results for
Planck, and is by far the simplest method of accounting for CMB
lensing in a parameter analysis: i.e. just pretend the lensed field is
Gaussian, and use the theoretical lensed CMB power spectrum. However
at high resolutions and sensitivity this will not be correct, as the statistics of the
$\tilde{C}_l$ on small scales are governed by the same small number of large scale
lensing modes~\cite{Hu:2001fa}.
 The naive approach could
be improved by (for example) calculating an empirical
$\hat{\tilde{C}}_l$ covariance from simulations and using this to make some
partial correction for non-Gaussianity induced variations to the
posterior distribution.  Non-Gaussian corrections should be small at Planck
sensitivity, however the non-Gaussian corrections to futuristic signal dominated lensing
$B$-mode observations can be important~\cite{Smith:2004up}. We have
checked that the lensed estimator $\hat{\tilde{C}}_l{}^{TT}$ and $\hat{\tilde{C}}_l{}^{EE}$
variances from simulations agree with the Gaussian results to within a
few percent at $l < 2000$.

\subsection{Sampling from the posterior}

We assume a simple flat adiabatic $\Lambda CDM$ cosmological model with the following parameters to
be determined from the data: primordial curvature
perturbation power spectrum with spectral index
$n_s$ and  amplitude $A_s$ (at wavenumber $0.05 \Mpc^{-1}$), baryon density $\Omega_b h^2$,
cold dark matter density $\Omega_c h^2$, optical depth $\tau $
(reionization assumed sharp), and Hubble parameter today $H_0= 100h
\Hunit$. We approximate the neutrinos as massless and assume standard
general relativity.

We use the
\COSMOMC\footnote{\url{http://cosmologist.info/cosmomc}}~\cite{Lewis:2002ah} Markov Chain Monte Carlo (MCMC) package to sample from
the posterior distribution of the parameters given the observed (simulated) data.
To make the posterior
parameter distributions more
Gaussian we use $\theta_r$ as a base parameter (with
flat prior) instead of the Hubble parameter $H_0$. The derived parameter $\theta_r$ is defined as the (approximate) ratio of the sound horizon at last
scattering to the angular diameter distance~\cite{Kosowsky02}, a
non-linear function of the other parameters that is
very well constrained by the position of the acoustic peaks.
We also
transform to the amplitude parameter $\ln A_s$ (with a flat prior) that is
constrained very well in a linear combination with $\tau$ because $A_s
e^{-2\tau}$ determines the small scale amplitude of the $C_l$. We then
use the covariance matrix to transform to an uncorrelated set of parameters that
the MCMC proposal density can use to explore the posterior distribution efficiently.
Given an
approximate covariance matrix from previous runs chains converge in a
few hours. For the first run the covariance can be learned
dynamically as the chain evolves (discarding samples from the evolving
part of the chain), or one can use the Hessian at the best fit point
as a useful starting approximation.

Since the computation time for generating the lensed $\tilde{C}_l$ is
dominated by the time to compute the transfer functions for the
unlensed $C_l$ and the lensing potential, parameters like $\ln A_s$
and $n_s$ remain `fast' parameters~\cite{Lewis:2002ah}, in that changing them is quick as
long as the other parameters remain fixed. Thus methods to efficiently
exploit the difference between `fast' and `slow' parameters~\cite{cosmomc_notes} can still
be used to speed up MCMC runs even when CMB lensing is included.

We neglect $\tilde{C}_l^{BB}$ since it is noise dominated even at Planck
sensitivity and has almost no effect on parameter constraints, though
including it in the Gaussian approximation is trivial if desired.

\subsection{WMAP data}

The importance of the lensing effect depends on the amplitude of the
potential fluctuations along the line of sight. Larger amplitudes cause
more lensing. By itself the first year WMAP data~\cite{Spergel:2003cb}
constrains the amplitude rather poorly due to a degeneracy with the
optical depth and the absence of $C_l^{EE}$ data to give a good upper
limit. Large values of the optical depth $\tau \agt 0.5$ allowed by
the data correspond to models with large amplitudes in which the
lensing effect is quite significant, so strictly speaking the lensing
should be accounted for in analyses of the first year WMAP data
alone~\cite{Bridle:2003sa,Tegmark:2003ud}. In a totally free WMAP
first year parameter
analysis the lensing already has a noticeable effect. However the region of parameter space with
$\tau \agt 0.5$ is almost certainly disallowed by numerous other
sources of data and so should not be taken too seriously.

The second year data WMAP should include $C_l^{EE}$ and a better
measurement of the third peak,
restricting to a much smaller range of $\tau$ and $\Omega_b h^2$, and so making the effect of
lensing on the tails of the distribution less
significant. As a simple toy model consider a full sky observation
with isotropic Gaussian noise with variance $N_l^{TT} = N^{EE}_l/4=N^{BB}_l/4 =  0.03 \mu K^2$ and a
symmetric Gaussian beam of $13$ arcminutes full-width half-maximum and
neglect foregrounds. We find that neglecting the lensing effect on the
power spectrum leads to a posterior mean of $\Omega_b h^2$
about half a standard deviation lower than it should be for a typical realization. This is
easily understood: lensing smoothes out the third peak in the
temperature $C_l$, making it appear lower relative to the first and
second peaks by a couple of percent. A similar effect can be produced
without lensing by lowering the baryon density, so unless lensing is
modelled consistently there is a danger of confusion.

Future ground or balloon results that can resolve the higher
temperature $C_l$ peaks
will also be sensitive to lensing as the corrections become larger
than $5\%$ at $l \agt 1500$.
We conclude that very soon it will be important to include
lensing in parameter analyses to obtain accurate results from CMB observations.

\subsection{Planck-like simulation}

\begin{figure}
\begin{center}
\psfig{figure=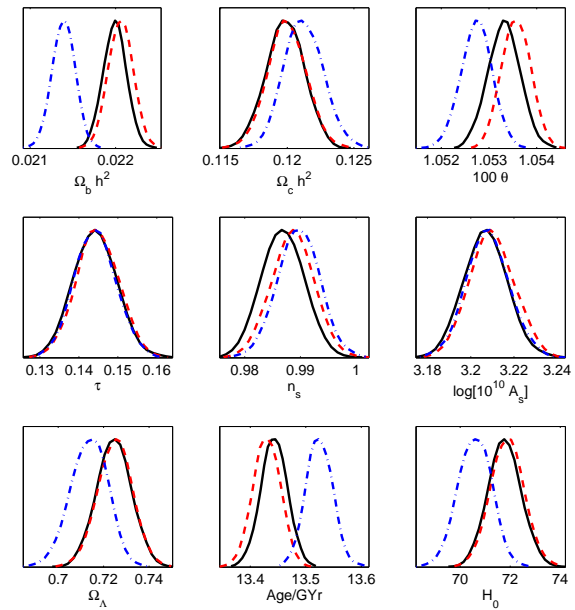,angle=0,width=7.5cm}
\caption{Parameter constraints from a simple Planck-like
  simulation. Solid lines analyse the lensed sky assuming Gaussianity
  with the lensed CMB power spectra, dashed lines are for an unlensed
  sky analysed with the unlensed power spectra, dash-dotted lines show
  the (inconsistent) result from analysing the lensed sky using the
  unlensed power spectra. The bottom row shows the dark energy
  density, age and Hubble constant derived parameters.
\label{planck_params}}
\end{center}
\end{figure}

As an example of a future observation where modelling the lensing will
be crucial we now consider a simple full-sky CMB observation
simulation at Planck-like sensitivity.
We compute theoretical CMB power spectra for the lensing potential and
unlensed fields for a simple fiducial model with $n_s =0.99$, $A_s = 2.5\times
10^{-9}$, $\Omega_b h^2 = 0.22$, $\Omega_c h^2 = 0.12$, $\tau = 0.15$, $h=0.72$, and simulate full sky
lensed maps with $12\times 2048^2\sim 5\times 10^7$ pixels (generated
by remapping a $12\times 8192^2\sim 8\times 10^8$ pixel unlensed sky
as described in the appendix). As a simple toy model of an
observation at optimistic Planck-like sensitivity we assume isotropic
Gaussian noise with variance
$N_l^{TT} = N^{EE}_l/4=N^{BB}_l/4 =  2\times 10^{-4} \mu K^2$ and a
symmetric Gaussian beam of $7$ arcminutes full-width half-maximum and
neglect foregrounds.

We use simulated lensed $\tilde{C}_l$ on scales with $l \le 2000$.
As discussed above we neglect non-Gaussianity of the lensed sky, but account
for the lensing by using accurate
theoretical lensed CMB power spectra~\cite{Challinor:2005jy}.
Obtaining parameter estimates
at Planck sensitivity is then no more difficult than with WMAP.

Fig.~\ref{planck_params} shows the posterior parameter constraints
from this particular realization. The constraints from the lensed sky
analysed using the lensed $\tilde{C}_l$ are in quite good agreement with
those one would have obtained if the unlensed sky were observable and
was analysed with the unlensed $C_l$. So current tools appear to
be sufficient to extract parameter constraints reliably at up to
Planck-like sensitivity.
Of course it is essential to model the lensing consistently: Fig.~\ref{planck_params}
also demonstrates the wrong parameter constraints that are obtained if the
lensed sky is incorrectly analysed using the unlensed $C_l$.

We appear to agree with
Ref.~\cite{Zaldarriaga:1997ch} that at this sensitivity lensing does
little to change the effectiveness of parameter estimation, with the
recovered error bars being of approximately the same width as when
analysing an unlensed sky. However we have not tested the accuracy of
the recovered error bars, merely showing that the posterior peaks are
at about the correct parameter values consistent with the error bar.
The result of Ref.~\cite{Hu:2001fa} would suggest that the
error bars are likely to be correct to $\alt 5\%$,
but for a
precision estimate of the error bars a fuller analysis would be
required. Increasing the error bars should make the lensed and unlensed
analysis results even more consistent, though there is no reason in general
why the posterior means should be identical in any given realization.

The effect of non-linear evolution of
the lensing potential can be safely neglected at Planck sensitivity.
However we have made one important linear-theory assumption, which is that the unlensed
CMB power spectra can be worked out accurately. Given a standard cosmology
and an ionization
history this is straightforward to do, however uncertainties in the complicated
details of recombination may mean that the ionization history is
not very well known\footnote{\url{http://cosmocoffee.info/viewtopic.php?t=174}}~\cite{Seager:1999km,Leung:2003je,Dubrovich:2005fc}. Future CMB
parameter analyses, including Planck,  may require a more
accurate calculation of recombination to obtain unlensed (and hence lensed)
power spectra accurately enough. In addition, other second order
signatures such as from the kinetic Sunyaev-Zel'dovich effect will
probably have a small but important contribution that also
needs to be accounted for~\cite{Knox:1998fp,Amblard:2004ih,Santos:2003jb,Zahn:2005fn}.

\section{Conclusions}

We have described a lensed CMB sky simulation code that can be used
for testing future analysis methods. As a simple benchmark we have
demonstrated parameter estimation at up to Planck sensitivity by using
the lensed CMB power spectrum on scales $l\le 2000$. This recovers the correct parameters,
though the error bars may be less accurate. More sensitive
observations will require a fuller analysis of the non-Gaussian
likelihood function. Higher resolution observations may require a
much more complicated numerical analysis of non-linear evolution of
the potential and
other non-linear effects~\cite{White:2003xz,Amblard:2004ih}.

Incidentally we have also demonstrated that
current parameter estimation methodology is sufficiently accurate for
parameter estimation with Planck (under the assumption of linearity
and assuming that the ionization history is known well enough). The simple lensing analysis
considered here only causes an order unity increase in computing time
compared to an unlensed analysis, so there is no problem accounting
for the lensing effect in parameter analyses.
Future CMB
parameter studies must account for CMB lensing to obtain correct
results.
The lensed CMB simulation
code is publicly
available\footnote{
Note that the latest version of LensPix uses a bicupic interpolation method described in the Appendix E.4 of Ref.~\cite{Hamimeche:2008ai}
\url{http://cosmologist.info/lenspix}}, as is the parameter estimation code including support for lensed CMB power spectra\footnote{\url{http://cosmologist.info/cosmomc}}~\cite{Lewis:2002ah}.

\section*{Acknowledgments}
I thank Matias Zaldarriaga, Anthony
Challinor, Carlo Contaldi, Ben Wandelt, Mike Nolta and Sarah Smith for
discussion and comments. Some of the results in this paper have been derived using a modified
version of the \Healpix~\cite{Gorski:2004by} package.
The beowulf computer used for some of this analysis was funded by the Canada
Foundation for Innovation and the Ontario Innovation Trust.

\appendix

\section{Harmonics and map making}

 We use spherical polar coordinates, with orthonormal basis
vectors at any point on the sphere given by
$\ve_\theta$ and $\ve_\phi$.  A complex  spin-$s$ quantity can be defined as
the components of a rank-$|s|$ tensor in the basis $\ve_\pm \equiv \ve_\theta \pm i\ve_\phi$.
For example the spin two polarization~\cite{Zaldarriaga97} is given in terms of the
polarization tensor $P_{ab}$ by ${}_2 P =\ve_+^a\ve_+^b P_{ab}$.
Further definitions and derivations using our notation can be found
in the appendix of Ref.~\cite{Lewis01}.

A spin $s$ quantity ${}_s \eta$ has harmonic components ${}_s
 a_{lm}$ given by
\begin{equation}
{}_s \eta = \sum_{lm} {}_s a_{lm} \,{}_s Y_{lm}
\end{equation}
where $l \ge |m|, l\ge |s|$. The spin harmonics are defined by
\begin{equation}
{}_sY_{lm} \equiv {}_s\lambda_{lm} e^{im\phi}\equiv \sqrt{\frac{(l-|s|)!}{(l+|s|)!}}\,\, \edth^{\!s}\, Y_{lm},
\label{eq:appbharm}
\end{equation}
and $\edth^{-|s|}\equiv (-1)^s \beth^{|s|}$, $Y_{lm}$ is a standard
spin zero harmonic and $\edth$ is the spin raising operator (see Ref.~\cite{Lewis01}). The harmonics have the symmetries
\begin{eqnarray}
{}_sY_{lm}^* &=& (-1)^{s+m} {}_{-s}Y_{l(-m)}\nonumber \\
{}_sY_{lm}(\pi-\theta,\phi)&=& (-1)^{l+m} {}_{-s}Y_{lm}(\theta,\phi).
\end{eqnarray}

We define gradient ($E$) and curl ($B$) harmonic components of a
tensor field with the general definition
\begin{eqnarray}
\sElm &\equiv& (-1)^H \half ({}_{|s|} a_{lm} + (-1)^s {}_{-|s|} a_{lm}) \nonumber\\
 i\,\sBlm &\equiv&  (-1)^H \half ({}_{|s|} a_{lm} - (-1)^s {}_{-|s|} a_{lm})
\end{eqnarray}
where $\sElm^* = (-1)^m \sE_{l(-m)}$, $\sBlm^* = (-1)^m \sB_{l(-m)}$
and $(-1)^H$ is a sign convention.
This definition ensures that gradient fields are always
pure $E$. In general a complex spin field
${}_{s} \eta$, with ${}_{|s|} \eta^* = {}_{-|s|}\eta$,  can be expanded as
\begin{eqnarray}
{}_{|s|} \eta &=&  (-1)^H\sum (\sElm + i\sBlm) {}_{|s|} Y_{lm} \nonumber\\
{}_{-|s|}\eta &=&  (-1)^{H+s}\sum_{lm} (\sElm - i\sBlm) {}_{-|s|} Y_{lm}.\,\,
\end{eqnarray}

For polarization ${}_2 P = \ve_+^a \ve_+^b P_{ab} = Q + iU$ where $Q$
and $U$ are the Stokes parameters measured in the $(\ve_\theta,
\ve_\phi)$ basis.
Where $Q$ and $U$ are instead measured with respect to a left handed set
$(\ve_\theta,-\ve_\phi)$ as in Refs~\cite{Kamionkowski97,Lewis01} we
have ${}_2 P = Q - iU$ (this basis defines a right handed set about
the incoming photon direction).
In \Healpix\footnote{\url{http://www.eso.org/science/healpix/}} 1.2 conventions ${}_2 P = Q + iU$, and $H=1$, whereas in
the conventions of Refs~\cite{Kamionkowski97,Lewis01} $H=0$. On
large scales $C_l^{TE} = (-1)^{H+1}|C_l^{TE}|$, and the $H=1$ convention
is used for the output from \CAMB\ and \CMBFAST\ so the large scale
correlation is positive. To be consistent with these codes we use the
$H=1$ convention for numerical work.

\subsection*{Harmonic transforms}

For CMB lensing we need to generate maps of the gradient of the
potential, a vector field on the sphere.  Various first order results (for example
the efficient quadratic estimators for the potential~\cite{Okamoto03})
also require transforming to and from spin one and spin three
fields.

To compute the spin $s$ harmonics one can either iterate the spin $s$
recursion relation (see Ref.~\cite{Lewis01}), or relate the harmonics
to the spin zero harmonics. Here we give the results for the latter
approach, which is followed by \Healpix~\cite{Gorski:2004by}.  Transforms of spin zero fields and spin two fields (for
polarization) are standard~\cite{Kamionkowski97} and included in
\Healpix\ 1.2.

Defining
\begin{multline}
 {}_{s} F^\pm_{lm}\equiv \sqrt\frac{(l-s)!}{(l+s)!} {}_s W^\pm_{lm} \equiv \half ({}_s\lambda_{lm} \pm (-1)^s{}_{-s}
 \lambda_{lm})\\
= \half ({}_s\lambda_{lm} \pm (-1)^m{}_{s}
 \lambda_{l(-m)})
\end{multline}
and using some results from Ref.~\cite{Lewis01} we have
\begin{eqnarray}
{}_1W^+_{lm} &=&  \frac{1}{\sin\theta}\left[
  \alpha_{lm} \lambda_{(l-1)m} - l \cos\theta\lambda_{l
    m}\right]\nonumber \\
{}_1 W^-_{lm} &=& \frac{m}{\sin\theta}\lambda_{lm}\nonumber\\
{}_2W^+_{lm} &=& \left[
    \frac{2(m^2-l)}{\sin^2\theta} - l(l-1)\right] \lambda_{lm}
      + \frac{2\cos\theta}{\sin^2\theta}\alpha_{lm}\lambda_{(l-1)m}\nonumber\\
{}_2W^-_{lm} &=&
  \biggr[  \alpha_{lm}\lambda_{(l-1)m} -
  (l-1)\cos\theta\lambda_{lm}\biggr]  \frac{2m}{\sin^2\theta}
  \nonumber\\
{}_3W^+_{lm} &=& \bigg[ \left(
    l(l-1)(l-2)- 4\frac{2l+m^2(l-3)}{\sin^2\theta}\right) \cos\theta \lambda_{lm} \nonumber\\
   &&- \alpha_{lm}
\left( l(l+1)+6 -\frac{4(2+m^2)}{\sin^2\theta}
    \right)\lambda_{(l-1)m}\bigg]\frac{1}{\sin\theta}\nonumber\\
{}_3W^-_{lm} &=& \biggl[ \left(
    4\frac{m^2-3l+2}{\sin^2\theta} -
    3(l-1)(l-2)\right)\lambda_{lm}\nonumber\\
 &&\qquad\qquad\qquad+12\alpha_{lm}
  \frac{\cos\theta\lambda_{(l-1)m}}{\sin^2\theta}\biggr] \frac{m}{\sin\theta},
\end{eqnarray}
where
\begin{equation}
\alpha_{lm} \equiv \sqrt\frac{(2l+1)(l^2-m^2)}{2l-1}.
\end{equation}
Similar results can be derived for higher spins if desired.
If several different spin transforms are being done at the same time
one can also use a relation like
\begin{multline}
l\sin\theta\, {}_{s+1} W_{lm}^\pm= (l-s)\left(m\,{}_s W^\mp_{lm} -
  l \cos\theta\, {}_s W_{lm}^\pm\right) \\
+ (l+s)\alpha_{lm}\, {}_s W^\pm_{(l-1)m}.
\end{multline}
To transform to and from a map of a spin field ${}_{|s|} \eta = R+iI$
(where $R$ and $I$ are real) we use the symmetry in $\theta$ so that
\begin{align}
&R(\theta,\phi) = (-1)^H \sum_{lm} \left( {}_sF^+_{lm} \sElm + i
  {}_sF^-_{lm} \sBlm\right)e^{im\phi} \nonumber\\
&I(\theta,\phi) =(-1)^H \sum_{lm} \left( {}_sF^+_{lm} \sBlm - i
  {}_sF^-_{lm} \sElm\right)e^{im\phi}\nonumber\\
&R(\pi-\theta,\phi) = (-1)^{H+s}\times\nonumber\\ & \hspace{1cm}\sum_{lm} (-1)^{l+m}\left( {}_sF^+_{lm} \sElm - i
  {}_sF^-_{lm} \sBlm\right)e^{im\phi} \nonumber\\
&I(\pi-\theta,\phi) = (-1)^{H+s} \times\nonumber\\ & \hspace{1cm}\sum_{lm} (-1)^{l+m}\left( {}_sF^+_{lm} \sBlm + i
  {}_sF^-_{lm} \sElm\right)e^{im\phi}.
\end{align}
For pixelizations where pixels are on
lines of constant latitude (such as \Healpix,
\IGLOO~\cite{Crittenden98} and \GLESP~\cite{Doroshkevich:2005xw}) the $\phi$ transform can be performed
rapidly using FFTs. The remaining computational cost is easily
parallelized by sending pixels at different $\theta$ to separate
processors, and $\lmax \approx 2000$, $n_{\text{pix}} \approx 10^7$ maps can
be generated in a few seconds over about fifty modern processors.

\subsection*{The deflection field}

The gradient of the potential (a scalar), $\grad \psi$, has harmonic components
\begin{equation}
{}_1E_{lm} = (-1)^{H+1}\sqrt{l(l+1)} \psi_{lm} \quad\quad {}_1 B_{lm} = 0
\end{equation}
which follows from
\begin{multline}
{}_1 a_{lm} = \int d\Omega (\ve_+\!\cdot \grad\psi)\, {}_1 Y_{lm}^* \\= - \int
d\Omega \edth \psi \,{}_1 Y_{lm}^*
=-\sqrt{l(l+1)} \psi_{lm}
\end{multline}
\begin{multline}
{}_{-1} a_{lm} = \int d\Omega (\ve_-\!\cdot \grad\psi)\, {}_{-1} Y_{lm}^* \\ =- \int
d\Omega \beth \psi\, {}_{-1} Y_{lm}^*
=\sqrt{l(l+1)} \psi_{lm}
\end{multline}
where  $\ve_\pm\!\cdot \grad\psi$
are the spin $\pm 1$ components of $\grad\psi$. Thus maps of the
gradient field can easily be constructed from the harmonic components
of $\psi$ using results from the last section. For spin one field
${}_1\eta = R+iI = \ve_+\cdot \vX$, we see that $R$
and $I$ are simply the $\ve_\theta$ and $\ve_\phi$ components
of the vector field $\vX$.

% Remember when visualizing that the Healpix
%convention is to view the sphere from the inside, so $\ve_\phi$ points
%left. Integrals over igloo pixels can be done analytically for spin $s$
%following the relations in my igloo notes.

\subsection*{Accurate lensed map making}
To make accurate lensed temperature maps on the full sky we use
\begin{equation}
 \tilde{T}(\vn) = T(\vn') = \sum_{lm} T_{lm} Y_{lm}(\vn').
\end{equation}
Given a set of harmonic coefficients, constructing the lensed map is
straightforward.  Using identities for spherical triangles,
the lensed temperature at a position $(\theta,\phi)$ is
given by the unlensed temperature at position $(\theta',\phi+\Delta\phi)$ where
\begin{eqnarray}
%\cos\theta' &=& \cos L\cos\theta + \sin L \sin\theta\cos\alpha\\
%\sin\Delta\phi &=& \frac{\sin\alpha \sin L}{\sin\theta'}
\cos\theta' &=& \cos d\cos\theta - \sin d \sin\theta\cos\alpha\\
\sin\Delta\phi &=& \frac{\sin\alpha \sin d}{\sin\theta'}
\end{eqnarray}
and the deflection vector is $\vd \equiv \grad\psi = d_\theta
\ve_\theta + d_\phi \ve_\phi = d\cos\alpha \ve_\theta + d\sin\alpha \ve_\phi$.
Except near the coordinate singularities the obvious
Euclidean results are a rather accurate approximation.

For the polarization the points move the same way, with the polarization
 maintaining the same orientation relative to the deflection
vector at the two points (neglecting field
rotation, see Ref.~\cite{Hirata:2003ka}). However for the components
of a spin field we have to be careful to account for the different
direction of the coordinate vectors at the two points~\cite{Challinor02}.
This requires rotating the components of the spherical polar coordinates
by $\gamma$, the difference between the angle made by $\ve_\theta$ and
the connecting geodesic at the two points. If $\alpha' \equiv
\alpha-\gamma$ we have
\begin{equation}
\tan(\alpha') = \frac{ d_\phi } { d \sin d \cot \theta +
  d_\theta \cos d }
\end{equation}
and after weak lensing a spin $s$ field ${}_s \eta$ becomes
\begin{equation}
{}_s \tilde{\eta}(\vn) = e^{is\gamma} {}_s \eta(\vn').
\end{equation}
For the spin two polarization field we can avoid inverse trigonometric
functions by using
\begin{equation}
e^{2i\gamma} = \frac{2 (d_\theta + d_\phi A)^2}{d^2(1+A^2)} -1 +
\frac{2i(d_\theta + d_\phi A)( d_\phi - d_\theta A) }{   d^2(1+A^2) }
\end{equation}
where $A\equiv \tan\alpha'$. Except near the poles this is very close
to unity.
%Indeed because the deflection angles are so small the
%rotation of the coordinate vectors between the two points is in
%general negligible, however we include it for completeness and to
%ensure the few pixels around the poles do not suffer any additional
%systematically worse errors than elsewhere.

Note that to get an accurate simulation of the lensing $B$ modes at $l
\agt 1000$ it is
necessary to include relatively high $l$ (more than just $\lmax +
500$ that is accurate for the other spectra). The $B$ lensing signal only
becomes useful for parameter estimation after Planck.

To generate the lensed field we can not use FFTs because even if $\vn$ is
sampled equally in $\phi$ on rings of constant $\theta$, the original positions $\vn'$
will not. An $\lmax \approx 2000$,
$n_{\text{pix}} \approx 10^7$ polarized map can be made in about 2000 CPU hours,
with good trivial parallelization.
In practice a much faster way to make maps to good accuracy is to
generate an unlensed sky at higher resolution and just re-map the
points. We
find that using $12\times 8192^2\sim 10^9$ pixels for the high resolution map
is sufficient to get a $12\times 1024^2$-pixel lensed sky with
$\tilde{C}_l$ accurate to $0.5\%$ at $l\alt 2000$. The fractional
accuracy on the polarization is similar, except for the $B$ modes
induced by lensing which are a more sensitive:  $12\times
16384^2\sim 3\times 10^9$
pixels are required to get $\tilde{C}_l^{BB}$ accurate at percent level
for $l\alt 2000$. The $B$-spectrum is quite sensitive to
$l_\text{max}$ and the non-linear power spectrum, so this accuracy
level is only notional. Note that choosing $l_\text{max}$ too low
generally underestimates the $B$-mode power, whereas pixelization
errors lead to an overestimation, so it is possible to get spuriously
 accurate power spectra without actually having computed the
lensed sky accurately.

The approximate point-remapping method is also
easily parallelized, and by only generating sections of the
high-resolution map on each cluster node, the total memory requirement
per node can remain below one gigabyte as long as enough nodes are
available. The scaling is then approximately the same as making maps at the higher
resolution, and even multi-billion pixel remappings can be done in under
an hour. For Planck resolution observations, lensed simulations can be
done in a few minutes with enough processors.
For applications where this speed is an issue some kind of faster
interpolation scheme might be useful (avoiding the need to generate a
fine map at much higher resolution than the base pixelization).

The parallelized code incidentally also provides a fast method for
performing spherical harmonic transforms (without lensing) on computer
clusters using MPI.

\bibliography{../../antony,../../cosmomc}

\end{document}